\documentclass[a4paper,11pt]{article}

\usepackage{amsthm,amsmath,amssymb,amsfonts,pstricks,pst-node,graphics}

\providecommand*{\pderiv}[3][]{%
        \frac{\partial^{#1}{#2}}%
                {\partial {#3}^{#1}}}

\def\ddt{\pderiv{}{t}}
\def\im{\operatorname{Im}}

\newcommand\A{{\mathcal A}}

\renewcommand\d{{\rm d}}

\newcommand\J{{\mathcal I}}
\newcommand\cH{{\mathcal H}}

\newcommand{\lieh}{{\mathfrak{h}}}

\def\maple{{\sl Maple}}

\newtheorem{Def}{Definition}
\newtheorem{The}{Theorem}
\newtheorem{Pro}{Proposition}

\newtheorem{Ex}{Example}
\newtheorem{Cor}{Corollary}

\begin{document}
\bibliographystyle{alpha}
\title{The Hunter-Saxton equation: remarkable structures of symmetries and conserved densities}
\author{Jing Ping Wang\\
School of Mathematics and Statistics, University of Kent, UK}
\date{}
\maketitle
\begin{abstract}
In this paper, we present extraordinary algebraic and geometrical structures for the
Hunter-Saxton equation: infinitely many commuting and non-commuting $x,t$-independent
higher order symmetries and conserved densities. Using a recursive relation, we
explicitly generate infinitely many higher order conserved densities dependent on
arbitrary parameters. We find three Nijenhuis recursion operators resulting from
Hamiltonian pairs, of which two are new. They generate three hierarchies of commuting
local symmetries. Finally, we give a local recursion operator depending on an
arbitrary parameter.

As a by-product, we classify all anti-symmetric operators of a definite form
that are compatible with the Hamiltonian operator $D_x^{-1}$.
\end{abstract}

\section{Introduction}
The Hunter-Saxton (HS) equation
\begin{equation}\label{HS}
 u_{xt}=2 u u_{xx}+u_x^2
\end{equation}
was proposed by Hunter and Saxton as an asymptotic model for the propagation of
weakly nonlinear unidirectional waves \cite{HS91}.  Its integrability was proved by
Hunter and Zheng \cite{HZ94} by studying the nonlocal evolution equation
\begin{equation}\label{HSn}
 u_{t}=2 u u_{x}-D_x^{-1} u_x^2,
\end{equation}
where $D_x^{-1}$ is the inverse of the total derivative \(D_x\). Indeed, equation
(\ref{HSn}) can be written
\begin{equation}\label{biHS}
u_t=D_x^{-1} \ \delta_u (- u u_x^2)=(u_x D_x^{-2} - D_x^{-2} u_x)\ \delta_u
(-u_x^2),
\end{equation}
where $\delta_u$ is variational derivative with respect to the dependent variable
$u$. Operators $D_x^{-1}$ and $u_x D_x^{-2} - D_x^{-2} u_x$ are Hamiltonian and form
a Hamiltonian pair. This leads to a recursion operator of the Hunter-Saxton equation,
\begin{eqnarray}\label{rec1}
 \Re=(u_x D_x^{-2} - D_x^{-2} u_x)D_x \ .
\end{eqnarray}
The $x$--derivative of Hunter-Saxton equation (\ref{HS}), that is
\begin{eqnarray}\label{xdhs}
u_{xxt}=2 u u_{xxx}+ 4 u_x u_{xx}
\end{eqnarray}
is closely related to the Camassa-Holm equation
\begin{eqnarray*}
u_t-u_{xxt}-3 u u_x+2 u_x u_{xx}+ u u_{xxx}=0\ .
\end{eqnarray*}
We often write the Camassa-Holm equation in the form:
$$m_t=u m_x+2 m u_x, \qquad m=u_{xx}-u \ .$$
Equation (\ref{xdhs}) corresponds to $m=u_{xx}$ under the time scaling transformation
$t\mapsto 2 t$. As pointed in \cite{HZ94}, the bi-Hamiltonian structure (\ref{biHS})
and the Lax pair of the Hunter-Saxton equation can be obtained from the corresponding
known structures of the Camassa-Holm equation. Geometrically, the Hunter-Saxton equation
(\ref{HS}) describes geodesic flow associated to the right-invariant metrics on a
homogeneous space \cite{khm03}. It is a particular case of the Euler-Poincar{\'e}
equation on the diffeomorphisms in one spatial dimension \cite{holm}.

Recently, equation (\ref{HSn}) was proposed as a model to describe shortwave
perturbation in a relaxing one-dimensional medium. Its integrability was studied by
introducing $v=D_x^{-1} u_x^2$. This leads to the hydrodynamic system
$$u_t=2 u u_x -v,\quad  v_t=2 u v_x,$$
which was called an integrable regularization of equation (\ref{HSn}). We refer to
\cite{Gole10,Sak09} and the references in them for the more details in this aspect.

In this paper, we look at the Hunter-Saxton equation (\ref{HS}) in its own right
instead of the traditional approach of relating it to the Korteweg-de Vries equation and the
Camassa-Holm equation \cite{khm03}. Quite surprisingly, we find it possesses
extraordinarily rich algebraic structures: possessing infinitely many
commutative and non-commutative $x,t$-independent higher order symmetries and
conservation laws. Besides the bi-Hamiltonian structure in (\ref{biHS}), we found
another two bi-Hamiltonian structures. Using these operators, we can obtain local
recursion operators for  the Hunter-Saxton equation (\ref{HS}).

We note that equation (\ref{xdhs}) can be linearised by the transformation
\cite{calo84}
\begin{eqnarray}\label{tran}
 v=\frac{u_x}{u_{xx}}, \qquad y=u_x\ .
\end{eqnarray}
The linearized equation is
\begin{eqnarray}\label{linhs}
v_t=-y^2 v_y-3 y v\ .
\end{eqnarray}
However, we didn't find a direct way to produce the results in this paper via the
linearisation. We also note that the calculation in this paper is purely algebraic.
We do not justify it in analytical sense.

The arrangement of the paper is as follows: In Section \ref{Sec2}, we define the
required concepts such as Hamiltonian, symplectic and Nijenhuis operators,
symmetries, cosymmetries, conservation laws and recursion operators for evolution
equations in the context of the variational complex. We then devote the rest of the
paper to the study of symmetries and conservation laws of the Hunter-Saxton equation
(\ref{HSn}). In Section \ref{Sec3}, we present a recursive relation to generate
infinitely many conserved densities (see Theorem \ref{th2}). We compute their Poisson brackets with
respect to the Hamiltonian operator $D_x^{-1}$ and obtain only three commuting pairs
among conserved densities $T_1^{(\alpha)}$ and  $T_2^{(\beta,\gamma)}$ defined in
Theorem \ref{th2}. In Section \ref{Sec4}, we find three recursion operators
corresponding to three commuting pairs obtained in the previous section (see Theorem \ref{th3}). To prove
that these operators are Nijenhuis, we classify all anti-symmetric operators of the
form
\begin{eqnarray*}
\cH=f(u_x,u_{xx}) D_x+D_x f +g(u_x,u_{xx},u_{xxx}) D_x^{-1} h(u_x,u_{xx},u_{xxx}) +h D_x^{-1} g,
\quad f\neq 0,
\end{eqnarray*}
which are compatible to Hamiltonian operator $D_x^{-1}$. Here $f$ is a smooth
function of $u_x$ and $u_{xx}$; $g$ and $h$ are smooth functions of $u_x$, $u_{xx}$ and
$u_{xxx}$. We list all five cases that arise (see Theorem \ref{th4}). From one of the Nijenhuis recursion operators, we
construct a parameter-dependent local recursion operator, which is no longer
Nijenhuis (see Corollary \ref{cor2}). Finally, we complete the paper with some discussion in Section \ref{Sec5}.

\section{Definitions}\label{Sec2}
In this section, we sketch the basic definitions of Hamiltonian, symplectic and
Nijenhuis operators following \cite{GD79, mr94j:58081, wang98}. In the context of the
variational complex we also define the some concepts for evolution equations such as
symmetries, cosymmetries, conservation laws and recursion operators, which also serves to fix our notation.
\subsection{Complex of variational calculus}
Let $x,t$ be the independent variables and $u$ be a  (vector-valued)
dependent variable. All smooth functions depending
on $u$ and $x$-derivatives of $u$ up to some finite, but unspecified order form a
differential ring $\A$ with total $x$-derivation
$$D_x=\sum_{k=0}^\infty u_{k+1}  \pderiv{}{u_k}, \quad \mbox{where}
\ u_k=\partial_x^k u. $$  The highest order of $x$-derivative we call the order of a
given function. For any element $g\in \A$, we define an equivalence class (or a
functional) $\int\! g$ by saying that $g$ and $h$ are equivalent denoted $g\equiv h$
if and only if \(g-h\in \im D_x\). Without causing confusion we sometimes write $g$
instead of $\int\! g$ . The space of functionals, denoted by $\A'$, does not inherit
the ring structure from $\A$.

The derivations on the ring $\A$ commuting with $D_x$ are known as evolutionary
vector fields. They are of the form $\partial_P=\sum_{k=0}^\infty D_x^k P
\pderiv{}{u_k} $.  Let $\lieh$ denote the space of all such $P$. The natural
commutator of derivations leads to the Lie bracket on $\lieh$, that is
\begin{eqnarray}\label{lie}
\begin{array}{c}[P, \ Q]=D_Q[P]-D_P[Q],\quad P, Q \in \eta,\end{array}
\end{eqnarray}
where $D_Q=\sum_{i=0}^\infty\frac{\partial Q}{\partial u_i} D_x^i$ is the
Fr{\'e}chet derivative of $Q$.

The action of any element $P\in \lieh$ on $\int g \in \A'$ can be defined as
\begin{eqnarray}\label{act}
\begin{array}{c}P \int g =\int \partial_P(g)=\int \sum_{k=0}^\infty D_x^k P
 \frac{\partial
g}{\partial u_k}=\int D_g[P].\end{array}
\end{eqnarray}

This action is a representation of the Lie algebra $\lieh$. We build up a Lie algebra
complex associated to it. This complex is called the complex of
variational calculus. Here we give the first few steps.

We denote the space of functional $n$-forms by $\Omega^n$ starting with
$\Omega^0=\A'$. We now consider the space $\Omega^1$.  For any vertical 1-form on the
ring $\A$, i.e., $\omega=\sum_{k=0}^\infty h^k  \d u_k$,
there is a natural non-degenerate pairing with an element $P\in \lieh$:
\begin{eqnarray}\label{pairing}
\begin{array}{c}<\omega, \ P>=\int \sum_{k=0}^\infty h^k \  D_x^k P =
\int \left(\sum_{k=0}^\infty (-D_x)^k h^k \right) \  P \ .\end{array}
\end{eqnarray}
Thus any element of $\Omega^1$ is completely defined by $\xi=\sum_{k=0}^\infty
(-D_x)^k h^k$.

The pairing (\ref{pairing}) allows us to give
the definition of (formal) adjoint operators to linear (pseudo)-differential
operators \cite{mr94g:58260}.
\begin{Def} Given a linear operator ${\cal S}: \lieh \rightarrow  \Omega^1$,
we call the operator
${\cal S}^{\star}: \lieh \rightarrow \Omega^1$ the adjoint operator of ${\cal S}$ if
$<{\cal S} P_1, \ P_2>=<{\cal S}^{\star} P_2, \ P_1>$, where $P_i\in \lieh$ for
$i=1,\ 2$.
\end{Def}
Similarly, we can define the adjoint operator for an operator mapping from $\Omega^1$
to $\lieh$, from $\lieh$ to $\lieh$ or from $\Omega^1$ to $\Omega^1$.

The variational derivative of each functional $ g\in \A'$ denoted by
 $\delta_u  g \in \Omega^1$ is defined so that
\begin{eqnarray}\label{euler}
\begin{array}{c}<\delta_u g,\ P>=(\d \int\! g) (P)=
<\sum_{k=0}^\infty (-D_x)^k \frac{\partial g}{\partial u_k},\ P> \ ,\end{array}
\end{eqnarray}
where $\d: \Omega^n\rightarrow \Omega^{n+1}$ is a coboundary operator. Due to the
non-degeneracy of the pairing (\ref{pairing}), we have $$\delta_u 
g=\sum_{k=0}^\infty (-D_x)^k \frac{\partial g}{\partial u_k}\in \Omega^1.$$ In the
literature one often uses ${\rm E}$ referring to the Euler
operator instead of $\delta_u$.

For any $\xi\in \Omega^1$, by direct calculation we obtain $\d \xi =
D_{\xi} - D_{\xi}^{\star}$. We say that the $1$-form $\xi$ is closed if $\d \xi=0$.

Finally, we give the formulas of Lie derivatives  along any $K\in \lieh$ using
Fr{\'e}chet derivatives, cf.  \cite{mr94j:58081} for the details.
\begin{Def}\label{def0}
Let $L_K$ denote the Lie derivative along $K\in \lieh$. We have
\begin{equation*}
\begin{array}{l} L_K g=\int\! D_{g}[K] \ \ \mbox{for} \ \  g\in\A';\\
L_K\! h=[K, h] \ \ \mbox{for} \ \ h\in\lieh;\\
L_K\!\xi =D_{\xi}[K] +D_K^{\star}(\xi) \ \ \mbox{for} \ \ \xi\in\Omega^1;\\
L_K\! \Re=D_{\Re}[K] -D_K \Re+\Re D_K \ \ \mbox{for} \ \ \Re: \lieh \rightarrow
\lieh;\\
L_K\! \cH=D_{\cH}[K] -D_K \cH-\cH D_K^{\star}\ \
\mbox{for} \ \  \cH: \Omega^1 \rightarrow \lieh;\\
L_K \!\J=D_{\J}[K] +D_K^{\star} \J+\J D_K\ \ \mbox{for} \ \  \J: \lieh \rightarrow
\Omega^1.\end{array}
\end{equation*}
\end{Def}
In this complex we can identify most of the important concepts in the study of integrable
systems such as symmetries, cosymmetries, conservation laws and recursion operators.
They are all characterised by the vanishing of the Lie derivatives with respect to a
given evolution equation. This will be discussed further in section \ref{sec23}.

\subsection{Symplectic, Hamiltonian and Nijenhuis operators}
\begin{Def} A linear operator ${\cal S}: \lieh \rightarrow \Omega^1$ (or
$ \Omega^1 \rightarrow  \lieh $) is anti-symmetric if $ {\cal S}= -{\cal S}^{\star}$.
\end{Def}

Given an anti-symmetric operator $\J: \lieh\rightarrow \Omega^1$, there is an
anti-symmetric $2$-form associated with it. Namely,
\begin{eqnarray}\label{2fom}
\begin{array}{c}\omega(P,Q)=<\J(P),Q>=-<\J(Q),P>=-\omega(Q,P), \quad P,Q\in \lieh .
\end{array}
\end{eqnarray}
Here the functional $2$-form $\omega$ has the canonical form \cite{mr94g:58260}
\begin{eqnarray}\label{2fomw}
\omega=\frac{1}{2} \int\!\! \d u \wedge \J \d u.
\end{eqnarray}
\begin{Def}
An operator $\J: \lieh\rightarrow \Omega^1$ is called symplectic if and only if the
anti-symmetric $2$-form (\ref{2fomw}) is closed, i.e., $\d\omega=0$.
\end{Def}
It is useful to know that for any $\xi\in \Omega^1$, if $\xi$ is not closed, the
operator $\d \xi= D_{\xi} - D_{\xi}^{\star}$ is a symplectic operator.  This can be
used to generate new symplectic operators or to determine whether a given operator is
symplectic or not.

For an anti-symmetric operator $\cH: \Omega^1\rightarrow \lieh$, we define a bracket
of two functionals $f$ and $g$ as
\begin{eqnarray}\label{poi}
\begin{array}{l}\left\{  f,\  g\right\}_{\cH}=<\delta_u f, \cH \delta_u g> .
\end{array}
\end{eqnarray}
\begin{Def}
The operator $\cH$ is Hamiltonian if the bracket defined by (\ref{poi}) is Poisson,
that is,  anti-symmetric and satisfies the Jacobi identity
$$\begin{array}{l}\left\{\left\{  f, g \right\}_{\cH},\
 h \right\}_{\cH}+\left\{\left\{  g,  h \right\}_{\cH},\
 f \right\}_{\cH}+\left\{\left\{  h,  f \right\}_{\cH},\
 g \right\}_{\cH}=0 .\end{array}$$
\end{Def}
For the Jacobi identity, there are several equivalent formulas given in
\cite{mr94j:58081} (see Theorem 5.1). In \cite{mr94g:58260} (see p. $443$), it is
formulated as the vanishing of the functional tri-vector:
$$ \int \theta \wedge D_{\cH} [\cH \theta] \wedge \theta =0 .$$
We are going to use it to classify all the Hamiltonian operators of a given family of
operators in Section \ref{Sec4}.

Let $\cH$ be a Hamiltonian operator. The Hamiltonian vector fields and their Hamiltonians
possess the property  \cite{mr94j:58081,mr94g:58260}:
\begin{eqnarray}\label{hamp}
\cH \delta_u\left\{ f,  g \right\}_{\cH}=[\cH \delta_u   f, \ \cH \delta_u g]\ .
\end{eqnarray}

The Jacobi identity is a quadratic relation for the operator $\cH$. In general, the
linear combination of two Hamiltonian operators is no longer Hamiltonian. If it is,
we say that these two Hamiltonian operators form a Hamiltonian pair. Hamiltonian pairs
play an important role in the theory of integrability. They naturally generate
Nijenhuis operators.
\begin{Def} A linear operator $\Re: \lieh\rightarrow \lieh$ is called a Nijenhuis
operator if it satisfies
\begin{eqnarray}\label{Nijen}
\begin{array}{l}[\Re P, \Re Q]-\Re[\Re P, Q]-\Re[P, \Re Q]+\Re^2[P, Q]=0,
\quad P, Q\in \lieh.\end{array}
\end{eqnarray}
\end{Def}

For a Hamiltonian pair $\cH_1$ and $\cH_2$, if $\cH_1$ is invertible, then operator
$\cH_2 \cH_1^{-1}$ is a Nijenhuis operator.

Using the definition of Lie bracket (\ref{lie}), formula (\ref{Nijen}) is equivalent
to
\begin{eqnarray}\label{Nijen1}
\begin{array}{l}L_{\Re P} \Re =\Re L_P \Re .\end{array}
\end{eqnarray}
The properties of Nijenhuis operators \cite{mr94j:58081} provide us with the
explanation how the infinitely many commuting symmetries and conservation laws of
integrable equations arise. In application, there are nonlocal terms in Nijenhuis
operators. A lot of work has been done to find sufficient conditions for Nijenhuis operators to
produce local objects \cite{ mr1974732,serg5, wang09}.
\subsection{Symmetries and conserved densities of evolution equations}\label{sec23}
To each element $K\in \lieh$, we associate an evolution equation of the form
\begin{eqnarray}\label{eq}
u_t=K.
\end{eqnarray}
Strictly speaking, one associates to the evolution equation the derivation
$$
\ddt+ \sum_{k=0}^\infty D_x^k K \pderiv{}{u_k}.
$$
As long as the objects concerned are explicitly time-independent as in this paper, there
is no difference. We refer to \cite{wang98,mr2002b:37100} for the case when objects
explicitly depend on time $t$.
\begin{Def}\label{defs}
Given an evolution equation (\ref{eq}), when the Lie derivatives of the following vanish
along $K\in \lieh$ we call: $ g\in\A'$ a conserved density; $h\in\lieh$  a
symmetry; $\xi\in\Omega^1$ a cosymmetry; $\Re: \lieh \rightarrow \lieh$ a recursion
operator; a Hamiltonian operator $\cH: \Omega^1 \rightarrow \lieh$ a Hamiltonian
operator for the equation; a symplectic operator $\J: \lieh \rightarrow \Omega^1 $ a
symplectic operator for the equation.
\end{Def}

From the above definitions, we can show that if $ f\in \A'$ is a conserved density
of the equation, then $\delta_u  f$ is its cosymmetry. Moreover, if $\cH$ is a
Hamiltonian operator and $\J$ is a symplectic operator of a given equation, then $\cH
\J$ is a recursion operator. The operator $\cH$ maps cosymmetries to symmetries while
$\J$ maps symmetries to cosymmetries.

We say that the evolution equation (\ref{eq}) is a Hamiltonian system if for a
(pseudo-differential) Hamiltonian operator $\cH$, there exists a functional $
f\in \A'$, called the Hamiltonian, such that $\cH\ \delta_u  f $ is a symmetry of the
equation. Additionally, if for a (pseudo-differential) symplectic operator $\J$,
which is compatible with $\cH$, there exists a functional $ g\in \A'$ such that
\begin{eqnarray*}
 \J u_t= \J K= \delta_u  g\ ,
\end{eqnarray*}
we say that the evolutionary equation is a (generalised) bi-Hamiltonian system.

\section{Conserved densities of the Hunter-Saxton equation}\label{Sec3}
In this section, we give the recursive relation to generate infinitely many
conserved densities for the Hunter-Saxton equation $T_k^{(\alpha_1,
\cdots,\alpha_k)}$, where $k$ is a nonnegative integer and $\alpha_i$ are parameters.
In general, these conserved densities are not in involution with respect to
the Hamiltonian operator $D_x^{-1}$. We show that there are only three commuting pairs
among $T_1^{(\alpha)}$ and  $T_2^{(\beta,\gamma)}$.
\begin{Pro}\label{pro1}
Equation (\ref{HS}) possesses a conserved density of the form $T=u_1^2
\left(\frac{u_2}{u_1^4}\right)^{\alpha}$ satisfying $D_t T=D_x( 2u T)$, where
$\alpha$ is a constant.
\end{Pro}
{\bf Proof.} According to the definition \ref{defs}, if $T$ is a conserved density,
then its Lie derivative along equation (\ref{HS}) vanishes. This is equivalent to
$D_t T \in \mbox{Im} D_x$. We have
\begin{eqnarray*}
&& D_t T=2 u_1 u_{xt}
 \left(\frac{u_2}{u_1^4}\right)^{\alpha}
 +\alpha  u_1^2 \left(\frac{u_2}{u_1^4}\right)^{\alpha-1}
\left( \frac{u_{xxt}}{u_1^4}-4 \frac{u_2 u_{xt}}{u_1^5} \right)\\
&=&2 u_1 (2 u u_2+u_1^2)  \left(\frac{u_2}{u_1^4}\right)^{\alpha}
+\alpha  u_1^2 \left(\frac{u_2}{u_1^4}\right)^{\alpha-1}
\left( \frac{2 u u_{xxx}}{u_1^4}-\frac{8 u u_2^2}{u_1^5} \right)
=D_x \left(2 u u_1^2 \left(\frac{u_2}{u_1^4} \right)^{\alpha} \right)
\end{eqnarray*}
and thus we proved the statement. \hfill $\diamond$

Since the Hunter-Saxton equation is free of any parameter, we get new conservation
laws by differentiating $T$ with respect to parameter $\alpha$. Thus we have
\begin{Cor}\label{cor1}
Expression $S=u_1^2 \left(\frac{u_2}{u_1^4}\right)^{\alpha} \left(\ln
\frac{u_2}{u_1^4} \right)^n$, $n\in \mathbb{N}$, is a conserved density of equation
(\ref{HS}) satisfying $D_t S=D_x(2 u S).$
\end{Cor}
In what follows we show how to build up more conserved densities using the above Proposition
and Corollary. First we prove the following general result.
\begin{The}\label{conlaw}
Assume that $F$ is a conserved density of equation (\ref{HS}) satisfying $F_t=D_x (2
u F)$. If for all solutions of equation (\ref{HS}) there exists a function $r$ such
that $r_t =2 u r_x$, then both $F r^{\alpha}$ and $G=F \left(\frac{D_x r}{u_1^2}
\right)^{\alpha}$ are conserved densities satisfying $D_t (F r^{\alpha})=D_x (2 u F
r^{\alpha})$ and $D_t G=D_x (2 u G)$ for any constant $\alpha$.
\end{The}
{\bf Proof.} We first check that $D_t (F r^{\alpha})$ is in the image of $D_x$.  Indeed
\begin{eqnarray*}
&&D_t (F r^{\alpha})=F_t r^{\alpha}+ \alpha F r^{\alpha-1} r_t
=D_x( 2u F) r^{\alpha}+ 2 u \alpha F r^{\alpha-1} r_x =D_x (2 u F r^{\alpha})\ .
\end{eqnarray*}
Next we show that
\begin{eqnarray*}
&&D_t\left( \left(\frac{D_x r}{u_1^2} \right)^{\alpha}\right)= \alpha \left(\frac{D_x
r}{u_1^2} \right)^{\alpha-1}
 \left(\frac{D_x r_t}{u_1^2}-2 \frac{ u_{xt} D_x r}{u_1^3}\right)\\
&=& \alpha \left(\frac{D_x r}{u_1^2} \right)^{\alpha-1}
\left(\frac{D_x (2 u r_x)}{u_1^2}-2 \frac{ (2 u u_2+u_1^2) D_x r}{u_1^3}\right)\\
&=& 2 \alpha  u \left(\frac{D_x r}{u_1^2} \right)^{\alpha-1}
\left(\frac{ r_{xx}}{u_1^2}-2 \frac{ u_2 D_x r}{u_1^3}\right)
=2 u  D_x \left(\frac{D_x r}{u_1^2} \right)^{\alpha} \ .
\end{eqnarray*}
Using the first part of the proof, we obtain that $G$ is a conserved density for any
constant $\alpha$. \hfill $\diamond$

We now search for function $r$ such that $r_t=2 u r_x$ for equation (\ref{HS}).
Notice that $r$ satisfies $r_t=2 u r_x$ if and only if  $\ln r$ satisfies the same
relation, that is, $(\ln r)_t=2 u (\ln r)_x$.
\begin{Pro}\label{pro2}
Assume that both $T$ and $T r$ are conserved densities of equation (\ref{HS})
satisfying $T_t=D_x (2 u T)$ and $D_t \left(T  r\right)=D_x (2 u T r)$. Then $r_t= 2
u r_x$ .
\end{Pro}
{\bf Proof.} From the assumption we have
\begin{eqnarray*}
0&=&D_t \left(T  r\right)-D_x (2 u T  r)=T_t  r+T r_t- D_x (2 u T)  r-2 u T r_x =T
\left(r_t-2 u r_x \right).
\end{eqnarray*}
This leads to the conclusion that $r_t=2 u r_x$. \hfill $\diamond$

From Proposition \ref{pro1} and Corollary \ref{cor1}, it follows that $r= \ln
\frac{u_2}{u_1^4}$ satisfies $r_t=2 u r_x$ and so does $r=\frac{u_2}{u_1^4}$. Using
Theorem \ref{conlaw}, we can now obtain the following result:
\begin{The}\label{th2}
The functionals $T_k^{(\alpha_1, \cdots,\alpha_k)}$ generated by the recursive relation:
\begin{eqnarray}
&&T_0=u_1^2 , \hspace{5.4cm}  r_0=-\frac{1}{u_1};\label{t0}\\
&&T_1^{(\alpha_1)}=T_0 r_1^{\alpha_1} ,  \hspace{4.4cm} r_1=\frac{1}{u_1^2} D_x r_0;
\label{t1}\\
&&T_2^{(\alpha_1,\alpha_2)}=T_1^{(\alpha_1)} r_2^{\alpha_2},\hspace{3.4cm}
r_2=\frac{1}{u_1^2} D_x r_1;\label{t2}\\
&& \cdots \cdots\nonumber\\
&&T_k^{(\alpha_1, \cdots,\alpha_k)}=T_{k-1}^{(\alpha_1, \cdots,\alpha_{k-1})}
r_k^{\alpha_k},\hspace{1.8cm} r_k=\frac{1}{u_1^2}
D_x r_{k-1};\label{tk}\\
&& \cdots \cdots\nonumber
\end{eqnarray}
are $(k+1)^{\rm th}$ order conserved densities of equation (\ref{HS}). Moreover,
$T_1^{(1)}\equiv 0$, $T_2^{(\alpha_1,1)}\equiv 0$ and when $\alpha_{k}\neq -1$ and
$k\geq 2$, we have
\begin{eqnarray}\label{equiv}
T_{k+1}^{(\alpha_1, \cdots,\alpha_{k},1)}\equiv -\sum_{i=1}^{k-1} \frac{\alpha_i}{\alpha_k+1}
T_k^{(\alpha_1, \cdots, \alpha_{i-1},
\alpha_{i}-1, \alpha_{i+1}+1, \cdots, \alpha_{k-1}, \alpha_{k}+1+\delta_{i,k-1} ) }  \ .
\end{eqnarray}
\end{The}
{\bf Proof}. We only need to prove the second part of the statement. It is easy to see that
 $$T_1^{(1)}=D_x r_0\equiv 0, \qquad T_2^{(\alpha_1,1)}=r_1 ^{\alpha_1} D_x r_1 \equiv 0.$$
We now show formula  (\ref{equiv}) by direct calculation:
\begin{eqnarray*}
&&T_{k+1}^{(\alpha_1, \cdots,\alpha_{k},1)}=T_{k}^{(\alpha_1, \cdots,\alpha_{k})} r_{k+1}
=r_1^{\alpha_1} r_2^{\alpha_2} \cdots r_{k}^{\alpha_{k}} D_x r_{k}\\
&\equiv&-\sum_{i=1}^{k-1} \frac{\alpha_i}{\alpha_k+1}r_1^{\alpha_1} \cdots r_{i-1}^{\alpha_{i-1}}
r_{i}^{\alpha_{i}-1} (D_x r_i) r_{i+1}^{\alpha_{i+1}}\cdots r_{k-1}^{\alpha_{k-1}}r_{k}^{\alpha_{k}+1}\\
&=&-\sum_{i=1}^{k-1} \frac{\alpha_i}{\alpha_k+1} u_1^2 r_1^{\alpha_1} \cdots r_{i-1}^{\alpha_{i-1}}
r_{i}^{\alpha_{i}-1} \frac{D_x r_i}{u_1^2} r_{i+1}^{\alpha_{i+1}}
\cdots r_{k-1}^{\alpha_{k-1}}r_{k}^{\alpha_{k}+1} \ .
\end{eqnarray*}
Thus we obtain the formula using the relation $r_{i+1}=\frac{D_x r_i}{u_1^2}$ and the
recursive relation for $T_k$. \hfill $\diamond$

For each $k\in \mathbb{N}$, the above $T_k^{(\alpha_1, \cdots,\alpha_k)}$ is
a $(k+1)^{\rm th}$ order conserved density with $k$ parameters.  In a similar way,
we can build up more conserved densities by adding more logarithms in front of $r_j$.
For instance, both $T_1^{\alpha_1} \left(\ln r_1 \right)^{\beta_1}$ and
$T_1^{\alpha_1} \left(\ln r_1 \right)^{\beta_1} \left( \ln \ln r_1 \right)^{\gamma_1}
$ are conserved densities.

Here all $T_k$ are local. By no means are the conserved densities we constructed above
complete. For example,  we haven't include the conserved density $\frac{1}{u_1}$,
that is,
$$D_t(\frac{1}{u_1})=D_x(\frac{2 u}{u_1}-3 x).$$
Besides, there are also nonlocal conserved densities. In \cite{HZ94},
the authors listed some of them generated by the recursion operator (\ref{rec1}).
For example, the conserved density $2 u^2 u_1^2+(D_x^{-1} u_1^2)^2$ depends on the same
nonlocal term $D_x^{-1} u_1^2$ as in equation (\ref{HSn}).

We can define the Poisson bracket of any two conserved densities with respect to a
Hamiltonian operator $D_x^{-1}$ according to formula (\ref{poi}):
\begin{eqnarray}\label{poid}
\left\{ T_i^{(\alpha_1, \cdots,\alpha_i)},\  T_j^{(\beta_1,
\cdots,\beta_j)}\right\}_{D_x^{-1}}
=<\delta_u ( T_i^{(\alpha_1, \cdots,\alpha_i)}),\ \  D_x^{-1} \delta_u( T_j^{(\beta_1,
\cdots,\beta_j)})> .
\end{eqnarray}

It is clear that
\begin{eqnarray}\label{poi12}
&&\left\{ T_0,\  T_j^{(\alpha_1, \cdots,\alpha_j)}\right\}_{D_x^{-1}}= <-2 u_2,\ \
D_x^{-1}\delta_u( T_j^{(\alpha_1, \cdots,\alpha_j)})>=<2 u_1,\ \
\delta_u( T_j^{(\alpha_1, \cdots,\alpha_j)})>\nonumber\\
&&= <2 u_1,\ \ D_{T_j^{(\alpha_1, \cdots,\alpha_j)}}^{\star}(1))>
=\int 2 D_x T_j^{(\alpha_1,
\cdots,\alpha_j)} =0\ .
\end{eqnarray}

With the help of the computer algebra system \maple\ , we obtain that
\begin{eqnarray}
&& \left\{ T_1^{(\alpha)},\  T_1^{(\beta)}\right\}_{D_x^{-1}}
=\frac{1}{2} \alpha \beta (\alpha-1)  (\beta-1) (\alpha-\beta)\ T_2^{(\alpha+\beta-5, 3)};
\label{t11}\\
&&\left\{ T_1^{(\alpha)},\  T_2^{(\beta,\gamma)}\right\}_{D_x^{-1}} =\frac{1}{2} \alpha \gamma
(\alpha-1)  (\gamma-1) (2-\gamma) \ T_3^{(\alpha+\beta-2,\gamma-3, 3)}\nonumber\\
&&\qquad \qquad+\frac{1}{2} \alpha \gamma (\alpha-1) (\gamma-1) (5 \alpha-3
\beta-10)\ T_3^{(\alpha+\beta-3,\gamma-1, 2)}\nonumber\\
&&\qquad \qquad+ \alpha  (\alpha-1) (\gamma-1) (6 \gamma+\gamma \alpha^2+\beta
-\alpha \beta \gamma+2 \beta \gamma-\beta^2-5 \gamma \alpha)
\ T_3^{(\alpha+\beta-4,\gamma+1, 1)}\nonumber\\
&&\qquad \qquad+ \alpha  \beta  (\alpha-1) (\alpha-2)  (1-\beta) (\gamma-1)\
T_2^{(\alpha+\beta-5,\gamma+3)}\ . \label{t12}
\end{eqnarray}
This implies that nontrivial $T_1^{(\alpha)}$ (i.e. $\alpha (\alpha-1)\neq 0$)
and $T_1^{(\beta)}$ do not commute unless $\alpha=\beta$.
Now we look at when a nontrivial $T_1^{(\alpha)}$ commutes with a
nontrivial $T_2^{(\beta,\gamma)}$. From Theorem \ref{th2},
we know that $T_3^{(\alpha_1,\alpha_2,1)}\equiv
\frac{-\alpha_1}{\alpha_2+1}T_2^{(\alpha_1-1,\alpha_2+2)}\ .$
Hence, the right-hand side of formula (\ref{t12}) vanishes if and only if
\begin{eqnarray*}
&& \left\{ \begin{array}{l} 2-\gamma=0;\\  5 \alpha-3 \beta-10=0;\\
-\frac{\alpha+\beta-4}{\gamma+2}(6 \gamma+\gamma \alpha^2+\beta -\alpha \beta
\gamma+2 \beta \gamma-\beta^2-5 \gamma \alpha)+(\alpha-2) \beta  (1-\beta)=0 \
.\end{array}\right.
\end{eqnarray*}
Solving this algebraic system, we obtain the following three solutions
\begin{eqnarray*}
 \left\{ \begin{array}{l}\alpha=\frac{1}{2} \\ \beta=-\frac{5}{2}\\ \gamma=2
        \end{array}\right. \qquad
 \left\{ \begin{array}{l}\alpha=-1 \\ \beta=-5\\ \gamma=2
        \end{array}\right. \qquad
\left\{ \begin{array}{l}\alpha=2\\ \beta=0\\ \gamma=2
        \end{array}\right.
\end{eqnarray*}
These lead to the following statement.
\begin{Pro}\label{pro3}
Assume that $\alpha(\alpha-1) \gamma (\gamma-1)\neq 0$. There are only three
commuting pairs among $T_1^{(\alpha)}$ and  $T_2^{(\beta,\gamma)}$, namely,
\begin{eqnarray*}
\left\{ u_2^{1/2}, \frac{(u_1 u_3-4 u_2^2)^2}{u_1^2 u_2^{5/2}}\right\}_{D_x^{-1}}
=\left\{ \frac{u_1^6}{u_2},  \frac{u_1^8 (u_1 u_3-4 u_2^2)^2}{ u_2^{5}}\right\}_{D_x^{-1}}
=\left\{ \frac{u_2^2}{u_1^6}, \frac{(u_1 u_3-4 u_2^2)^2}{u_1^{12}}\right\}_{D_x^{-1}}
=0\ .
\end{eqnarray*}
\end{Pro}
Note that the Poisson bracket of any pair of  conserved densities is again a conserved
density since
\begin{eqnarray*}
&&L_{K} \left\{ T_i^{(\alpha_1, \cdots,\alpha_i)},\  T_j^{(\beta_1,
\cdots,\beta_j)}\right\}_{D_x^{-1}}\\
&=& \left\{L_K \left( T_i^{(\alpha_1, \cdots,\alpha_i)}\right),\  T_j^{(\beta_1,
\cdots,\beta_j)}\right\}_{D_x^{-1}} + \left\{ T_i^{(\alpha_1, \cdots,\alpha_i)},\  L_{K}
\left(T_j^{(\beta_1, \cdots,\beta_j)}\right)\right\}_{D_x^{-1}}=0.
\end{eqnarray*}
It is surprising to see the results are linear combinations of the conserved
densities listed in Theorem \ref{th2} since they are by no means complete. It seems
that these conserved densities are closed under the defined Poisson bracket (\ref{poid}).

In next section, we show that the above three commuting pairs lie in three different
commuting hierarchies generated by three Nijenhuis recursion operators. The natural
question is whether there are more commuting pairs if we compute the Poisson bracket
between higher order conserved densities. In Appendix A, we include the formula for
$\left\{ T_1^{(\alpha)},\  T_3^{(\beta,\gamma,\mu)}\right\}_{D_x^{-1}} $. Based on
it, we find three commuting pairs between conserved densities of second order and
those of fourth order, generated by the same three Nijenhuis operators. We conjecture
that there are only three commuting hierarchies starting with $ T_1^{(\alpha)}$. By
computing the Poisson bracket between $ T_2^{(\alpha_1, \alpha_2)}$ and $
T_3^{(\beta_1,\beta_2,\beta_3)}$, we discover another commuting pair. It is listed in
Appendix A. However, we have not found the corresponding Nijenhuis recursion
operator.

\section{Symmetries and Recursion operators}\label{Sec4}
We know from (\ref{biHS}) that the operator $D_x^{-1}$ is a Hamiltonian operator for
equation (\ref{HSn}), mapping cosymmetries (the variational derivatives of conserved
densities) to symmetries. Thus, we can produce infinitely many symmetries from the
conserved densities listed in Theorem \ref{th2}. Using the property of Hamiltonian
operators (\ref{hamp}), we have
\begin{eqnarray}\label{hampro}
D_x^{-1} \delta_u\left\{ T_i^{(\alpha_1, \cdots,\alpha_i)},\
T_j^{(\beta_1, \cdots,\beta_j)}\right\}_{D_x^{-1}}
=[D_x^{-1} \delta_u ( T_i^{(\alpha_1, \cdots,\alpha_i)}),\ \
D_x^{-1} \delta_u( T_j^{(\beta_1,
\cdots,\beta_j)})].
\end{eqnarray}
From the results in the previous section, we know some symmetries are commuting and some
are not, cf. formula (\ref{t11}) and (\ref{t12}). In this section, we will present
some recursion operators to generate infinitely many commuting and noncommuting
symmetries of the Hunter-Saxton equation (\ref{HSn}).
\subsection{Nijenhuis recursion operators and commuting symmetries}\label{sec41}
The Nijenhuis recursion operators \cite{olv77} are used to generate infinitely many
local commuting symmetries for integrable equations. In this section, we present
three Nijenhuis recursion operators for the Hunter-Saxton equation (\ref{HSn}), which
correspond to three commuting pairs in Proposition \ref{pro3}. We then prove that
they generate infinitely many local symmetries.
\begin{The}\label{th3} The following three operators:
\begin{eqnarray}
&&\Re_1=\left( 2 u_2^{-1} D_x +2 D_x u_2^{-1} - u_3 u_2^{-3/2} D_x^{-1} u_3
u_2^{-3/2} \right) D_x\label{r1} \\
&&\Re_2= \left(  u_1^4 u_2^{-2} D_x + D_x u_1^4 u_2^{-2} -8 u_1 D_x^{-1} u_1
\right) D_x \label{r2}\\
&&\Re_3= \left(u_1^{-4} D_x\! +\! D_x u_1^{-4}\!\! -4(u_3 u_1^{-6}\! \!
-3 u_2^2 u_1^{-7}) D_x^{-1}\! u_1
\! -4 u_1 D_x^{-1}\! (u_3 u_1^{-6}\!-3 u_2^2 u_1^{-7}) \right) \!D_x \label{r3}
\end{eqnarray}
are all recursion operators of equation (\ref{HSn}).
\end{The}
{\bf Proof}. According to Definition \ref{defs}, we check whether $L_K\! \Re_i$
vanishes for each $i=1,2,3$, where $K=2 u u_1 -D_x^{-1} u_1^2.$  Since the
calculation is similar, we only work it out for $i=2$. To simplify the computation,
we introduce some notation: $s= u_1^4 u_2^{-2}$ and  $s_j=D_x^j s$. So
\begin{eqnarray*}
&&D_K \Re_2-\Re_2 D_K\\
&=& 2 (u D_x + u_1- D_x^{-1} u_1 D_x) ( s D_x + D_x s -8 u_1 D_x^{-1} u_1) D_x\\
&&-2 ( s D_x + D_x s -8 u_1 D_x^{-1} u_1) D_x (u D_x + u_1- D_x^{-1} u_1 D_x)\\
&=& 2 (u D_x + D_x^{-1} u_2) ( s D_x + D_x s -8 u_1 D_x^{-1} u_1) D_x\\
&&-2 ( s D_x + D_x s -8 u_1 D_x^{-1} u_1) (u D_x + u_1+u_2 D_x^{-1}) D_x\\
&=& \left( (2 u s_1-4 s u_1) D_x+ D_x (2 u s_1-4 s u_1) -
8(2 u u_2 + u_1^2) D_x^{-1} u_1 +D_x^{-1}
(8 u_1^3 -4 s u_3-2 s_1 u_2)\right.\\
&&\left.- 8 u_1 D_x^{-1} (2 u u_2 + u_1^2) +(8 u_1^3  -4  s u_3-2 s_1 u_2 )
D_x^{-1}\right) D_x \\
&=& \left( 4 (2 u u_1^3 u_2^{-1}-u u_1^4 u_2^{-3} u_3-u_1^5 u_2^{-2}) D_x
+ 4 D_x (2 u u_1^3 u_2^{-1}-u u_1^4 u_2^{-3} u_3-u_1^5 u_2^{-2})\right.\\
&&\left.
- 8(2 u u_2 + u_1^2) D_x^{-1} u_1
- 8 u_1 D_x^{-1} (2 u u_2 + u_1^2)\right) D_x ,
\end{eqnarray*}
which equals to $D_{\Re_2}[K]$. Therefore,  we have $L_K\! \Re_2=0$ and thus 
the statement is proved. \hfill $\diamond$

The recursion operator (\ref{rec1}) given in \cite{HZ94} is the inverse operator of
$\Re_1$. Indeed, we can prove the following statement:
\begin{Pro}
 $\Re_1=8 \Re^{-1}=8 D_x^{-1} (u_1D_x^{-2}-D_x^{-2} u_1)^{-1}$.
\end{Pro}
{\bf Proof}. Note that $u_1D_x^{-2}-D_x^{-2} u_1= D_x^{-1} (u_2 D_x^{-1} +D_x^{-1}
u_2) D_x^{-1}.$ To prove the statement, we only need to show that
 $8 \left(u_{2}D_x^{-1}+D_x^{-1}u_{2}\right)^{-1}=2 u_2^{-1}
D_x +2 D_x u_2^{-1} - u_3 u_2^{-3/2} D_x^{-1} u_3
u_2^{-3/2}$. Indeed,
\begin{eqnarray*}
&&(2 u_2^{-1} D_x +2 D_x u_2^{-1} - u_3 u_2^{-3/2} D_x^{-1} u_3
u_2^{-3/2}) (u_{2}D_x^{-1}+D_x^{-1}u_{2})\\
&=& 8\! +\! 2 u_2^{-1} u_3 D_x^{-1}\!\! -\!\!2 u_2^{-2} u_3 D_x^{-1} u_2
\!\!-\!\!2 u_3 u_2^{-3/2} D_x^{-1} (u_2^{1/2})_x D_x^{-1}+ 2 u_3 u_2^{-3/2}
D_x^{-1} (u_2^{-1/2})_x D_x^{-1} u_2\\
&=& 8 + 2 u_2^{-1} u_3 D_x^{-1} -2 u_2^{-2} u_3 D_x^{-1} u_2
-2 u_3 u_2^{-1}  D_x^{-1}+2 u_3 u_2^{-3/2} D_x^{-1} u_2^{1/2}\\
&&+ 2 u_3 u_2^{-2}  D_x^{-1} u_2
- 2 u_3 u_2^{-3/2} D_x^{-1} u_2^{1/2}\\
&=&8,
\end{eqnarray*}
and this leads to the statement.
\hfill $\diamond$

In the paper \cite{HZ94}, the authors proved that the recursion operator (\ref{rec1})
is the ratio of a Hamiltonian pair. Hence it is a Nijenhuis operator and so is its
inverse operator $\Re_1$ \cite{mr94j:58081}. We now prove that the other two
recursion operators $\Re_2$ and $\Re_3$ are also the ratio of Hamiltonian pairs. To
do so, we classify all anti-symmetric operators of the form
\begin{eqnarray}\label{Ham}
\cH=f(u_1,u_2) D_x+D_x f +g(u_1,u_2,u_3) D_x^{-1} h(u_1,u_2,u_3) +h D_x^{-1} g,
\quad f\neq 0,
\end{eqnarray}
which are compatible with the Hamiltonian operator $D_x^{-1}$. Here $f$ is a smooth
function of $u_1$ and $u_2$; $g$ and $h$ are smooth functions of $u_1$, $u_2$ and
$u_3$. We give the result below and the proof in Appendix B.
\begin{The}\label{th4}
If an anti-symmetric operator (\ref{Ham}) forms a Hamiltonian pair with the Hamiltonian
operator $D_x^{-1}$, then the smooth functions $f$, $g$ and $h$ are one of the
following five cases. Here $a,\ b, \ c, \ d$ are smooth functions of $u_1$ and
$c_i\in \mathbb{C}, \ i=1,2, \cdots ,5$ are constants.

\begin{eqnarray}
&&{\rm I}. \qquad\ \left\{\begin{array}{l} f=\frac{1}{a(u_1)^2 u_2^2},
\quad a(u_1)\neq 0\\h=b(u_1)\\ g=c(u_1) \end{array}\right.\label{case1}\\ 
&&{\rm II}. \quad\ \ \left\{\begin{array}{l} f=\frac{1}{(a(u_1) u_2+c_1)^2}\\h=c_2 u_1 +c_3\\
g=c_4 u_1+c_5\end{array}\right.\label{case2}\\ 
&&{\rm III}. \quad \left\{\begin{array}{l} f=\frac{4}{(a(u_1) u_2+b(u_1))^2},
\qquad a(u_1)\neq 0 \\h=c_1 u_1+c_2 \\
g= -\frac{16 u_3}{(a(u_1) u_2+b(u_1))^3}+ 8  \frac{\partial}{\partial
u_1} \left(\frac{b+2 a u_2}{a(u_1)^2 (a(u_1) u_2+b(u_1))^2}\right)
+c(u_1)\\
b(u_1)=c_1 u_1^2+2 c_2 u_1+c_3\\
c(u_1)=\frac{b' c_4+(u_1 b'-b)c_5}{b(u_1)^2} + \frac{16 a'b+8 a
b'}{a(u_1)^3 b(u_1)^2}
\end{array}\right.\label{case3}\\ 
&&{\rm IV}. \quad\ \left\{\begin{array}{l} f=\frac{4}{b(u_1)^2} \\h=c_1 u_1+c_2\\
g= -\frac{16 u_3}{b(u_1)^3}+\frac{24 b' u_2^2}{b^4}+c(u_1)\\
b(u_1)=c_1 u_1^2+2 c_2 u_1+c_3\\
c(u_1)=\frac{b' c_4+(u_1 b'-b)c_5}{b(u_1)^2}
\end{array}\right.\label{case3'}
\end{eqnarray}
\begin{eqnarray}
&&{\rm V}. \qquad \left\{\begin{array}{l} f=\frac{1}{a(u_1) u_2^2+b(u_1)u_2+c(u_1)}
\\h=\frac{c_1 u_3}{(a(u_1) u_2^2+b(u_1)u_2+c(u_1))^{3/2}}-\frac{1}
{2 c _1 c_2}\frac{\partial}{\partial u_1}
\left(\frac{b u_2+2 c}{(a(u_1) u_2^2+b(u_1)u_2+c(u_1))^{1/2}}\right)+d(u_1)
\\ g= c_2 h - 2 c_2 d(u_1)\\
d(u_1)=c_3 u_1+c_4\\
4 a c -b^2= 4 c_1^2 c_2\\
\frac{\partial^3 c(u_1)}{\partial u_1^3}= 6 c_1^2 c_2^2 c_3 d(u_1)
\end{array}\right.\label{case4}
\end{eqnarray}
\end{The}
\begin{Ex}
Two special cases from case II (cf. (\ref{case2})) lead to Hamiltonian pairs
$$D_x+2 u_1 D_x^{-1} +2 D_x^{-1} u_1 +\lambda D_x^{-1}$$
and $$D_x+2 u_1 D_x^{-1} u_1+\lambda D_x^{-1}\ .$$ These are the bi-Hamiltonian
structures for potential Korteweg-de Vries equation and potential modified
Korteweg-de Vries equation respectively since
$$u_t=u_3+3 u_1^2=D_x^{-1} \delta_u (\frac{1}{2}u_2^2-u_1^3)=
(D_x+2 u_1 D_x^{-1} +2 D_x^{-1} u_1)\delta_u (-\frac{1}{2} u_1^2)$$ and
$$u_t=u_3+u_1^3=D_x^{-1} \delta_u (\frac{1}{2}u_2^2-\frac{1}{4}u_1^4)=(D_x+2 u_1
D_x^{-1} u_1)\delta_u (-\frac{1}{2} u_1^2)\ .$$
\end{Ex}
The recursion operators (\ref{r2}) and (\ref{r3}) in Theorem \ref{th3} are the ratios
of Hamiltonian pairs listed in Theorem \ref{th4}. It is obvious that the recursion
operator (\ref{r2}) can be derived from Case I (\ref{case1}).

For Case IV, if we take  $b=2 u_1^2$, $c(u_1)=0$ and $h= 2 u_1$, then we get
$g=-2 u_1^{-6} u_3+6 u_1^{-7} u_2^2$. This leads to a Hamiltonian pair
$$u_1^{-4} D_x\! +\! D_x u_1^{-4}\!\! -4(u_3 u_1^{-6}\! \!
-3 u_2^2 u_1^{-7}) D_x^{-1}\! u_1 \! -4 u_1 D_x^{-1}\! (u_3 u_1^{-6}\!-3 u_2^2
u_1^{-7}) +\lambda D_x^{-1},$$ from which we obtain the recursion operator $\Re_3$
(\ref{r3}) in Theorem \ref{th3}. We can also directly get the recursion operator
(\ref{r1}) from Case V by taking $a(u_1)=c(u_1)=d(u_1)=0$ and $b(u_1)=\frac{1}{2}$.
Since all three recursion operators can be obtained from Hamiltonian pairs, they are
all Nijenhuis operators.
\begin{The}\label{th5}
The recursion operators (\ref{r1})--(\ref{r3}) are Nijenhuis operators.
\end{The}
We can also view the recursion operators (\ref{r1})--(\ref{r3}) as the products of
Hamiltonian and symplectic operators. In a recent paper \cite{wang09}, we proved that
for Nijenhuis operators that are the products of weakly nonlocal Hamiltonian and
symplectic operators \cite{MaN01}, hierarchies of commuting local symmetries and
conserved densities in involution can be generated under some easily verified
conditions. To be self-contained, we restate the result in  \cite{wang09} valid for
the operators in this paper:
\begin{quotation}
\noindent Consider a Hamiltonian operator $\cH$ of the form (\ref{Ham}) and a symplectic
operator $\J=D_x$ such that $\cH \J$ is a Nijenhuis operator. Assume that $L_{g}
\J=L_{g}\cH =L_{h} \J=L_{h}\cH =0$. If there exists a closed $1$-form $\xi$
satisfying $L_{g} \xi=L_h \xi =L_{\cH \xi} \xi=0$ such that $\J \cH \xi$ is closed,
then all $(\J\cH)^j \xi $ are closed $1$-forms and $\cH (\J\cH)^j \xi $
 commute for $j=0, 1, 2 , \cdots $.
\end{quotation}
We denote the recursion operators (\ref{r1})--(\ref{r3}) in Theorem \ref{th3} as
$\Re_i=\cH_i D_x$, $i=1,2,3$. Checking the conditions in this statement for all
$\Re_i$,  we can prove the following theorem.
\begin{The}\label{th6} Let $\xi_1=\delta_u (u_2^{1/2})$, $\xi_2
=\delta_u (\frac{u_1^6}{2 u_2})$ and $\xi_3=\delta_u (\frac{u_2^2}{2 u_1^6})$. Then
all $\Re_i^{\star j} \xi_i $ are closed $1$-forms and for each fixed $i \in \left\{
1,2,3\right\}$, all
 $\cH_i \Re_i^{\star j} \xi_i $ commute for $j=0, 1, 2 , \cdots $.
\end{The}
{\bf Proof}. We write out the proof for the recursion operator $\Re_2$. The proof for
operators $\Re_1$ and $\Re_3$ is similar, and will not repeat it. Since $\xi_2$ is the
variational derivative of $\frac{u_1^6}{2 u_2}$, it is clearly a closed $1$-form. In
this case, we have $g=h=u_1$ and trivially $L_{u_1} D_x=L_{u_1}\cH_2=L_{u_1} \xi_2
=0$. We only need to show that $L_{\cH \xi_2} \xi_2=0$ and  $D_x \cH_2 \xi_2$ is
closed. Note that
\begin{eqnarray*}
&&\xi_2=\frac{u_1^6 u_4}{u_2^3}-3 \frac{u_1^6 u_3^2}{u_2^4}+12 \frac{u_1^5
u_3}{u_2^2} -30 u_1^4;\\
&&\cH_2 \xi_2=2 \frac{u_1^{10} u_5}{u_2^5}-20 \frac{u_1^{10} u_3 u_4}{u_2^6}+40
\frac{u_1^9u_4}{u_2^4}+30 \frac{u_1^{10} u_3^3}{u_2^7}-120 \frac{u_1^9 u_3^2}{u_2^5}
+220 \frac{u_1^8 u_3}{u_2^3} -320 \frac{u_1^7}{u_2};
\end{eqnarray*}
and $D_x \cH_2 \xi_2=-\delta_u (\frac{u_1^8(u_1 u_3-4 u_2^2)^2}{u_2^5})$, which
implies that $D_x \cH_2 \xi_2$ is closed. Using the computer algebra system \maple\ ,
we can check $L_{\cH \xi_2} \xi_2=D_{\xi_2} [\cH \xi_2]+D_{\cH \xi_2}^{\star}
(\xi_2)=0$. Thus we prove the statement for $\Re_2$.  \hfill $\diamond$

We have proved that the recursion operators (\ref{r1})--(\ref{r3}) are Nijenhuis.
Using them, we generate three hierarchies of commuting symmetries. However, the
elements in the different hierarchies do not commute.

Note that the Lie derivative is a derivation. Thus the products and additions of
recursion operators are recursion operators. For instance, operators $\Re_1 \Re_2$
and $[\Re_1, \Re_3]$ are also recursion operators of the Hunter-Saxton equation
(\ref{HSn}). In general, they are no longer Nijenhuis (cf. (\ref{Nijen})) and do not
generate hierarchies of local symmetries.
\subsection{Recursion operators and noncommuting symmetries}\label{sec42}
In this section, we first prove that the adjoint operator of the recursion operator
$\Re_2$ (\ref{r2}) acting on any cosymmetry $\delta_u
(T_j^{(\alpha_1,\cdots,\alpha_j)})$ produces local cosymmetries. Since the resulting
cosymmetries are not closed, we can construct symplectic operators, which depend
on parameters. Further, this leads to parameter-dependent recursion operators.
\begin{Pro}\label{pro5}
Starting from any symmetry $Q_0=D_x^{-1} \delta_u (T_j^{(\alpha_1,
\cdots,\alpha_j)})$, where $T_j^{(\alpha_1, \cdots,\alpha_j)})$ are listed in Theorem
\ref{th2}, all $Q_k=\Re_2^k Q_0$ are local symmetries $k=0,1,2, \cdots$.
Equivalently, all $\Re_2^{\star k} \delta_u (T_j^{(\alpha_1, \cdots,\alpha_j)})$ are
local cosymmetries, where $\Re_2^{\star}$ is the adjoint of operator $\Re_2$.
\end{Pro}
{\bf Proof}. Note that $\Re_2=\left(u_1^4 u_2^{-2} D_x + D_x u_1^4 u_2^{-2} -8 u_1
D_x^{-1} u_1\right) D_x$. To prove that $Q_k$ are local, we only need to show  that
$u_2 Q_{k}$ is in the image of $D_x$. We prove the statement by induction.

We know that from (\ref{poi12}) the Poisson bracket of $T_0$ and $T_j^{(\alpha_1,
\cdots,\alpha_j)}$ vanishes. This implies that $u_2 Q_0 \in \mbox{Im} D_x$. Assume
that $u_2 Q_{k-1} \in \mbox{Im} D_x$. We now show that $u_2 Q_{k} \in \mbox{Im} D_x$.
Indeed,
\begin{eqnarray*}
u_2 Q_k&=&u_2 \Re_2 Q_{k-1}= u_2 \left(  u_1^4 u_2^{-2} D_x + D_x u_1^4 u_2^{-2} -8
u_1 D_x^{-1} u_1 \right) D_x Q_{k-1}\\
&=& 2 u_1^4 u_2^{-1} D_x^2 Q_{k-1}\! -2 u_1^4 u_2^{-2} u_3 D_x Q_{k-1}\!+ 4  u_1^3 D_x
Q_{k-1}\! -8 u_1^2 u_2 Q_{k-1}\!+ 8 u_1 u_2 D_x^{-1} u_2 Q_{k-1}\\
&=& D_x (2 u_1^4 u_2^{-1} D_x Q_{k-1}-4 u_1^3 Q_{k-1}+4 u_1^2 D_x^{-1} u_2 Q_{k-1} )
\in \mbox{Im} D_x\ .
\end{eqnarray*}
Thus all $Q_k$ are local. Note that
$$u_1 \Re_2^{\star k} \delta_u (T_j^{(\alpha_1,\cdots,\alpha_j)})
=u_1 D_x \cH_2\cdots D_x \cH_2  \delta_u (T_j^{(\alpha_1,\cdots,\alpha_j)})
=u_1 D_x Q_k=D_x (u_1 Q_k)-u_2 Q_k\ .$$
We have just proved that $u_2 Q_{k} \in \mbox{Im} D_x$. Thus we have
$u_1 \delta_u (T_j^{(\alpha_1, \cdots,\alpha_j)})\in \mbox{Im} D_x$.
Hence all $\Re_2^{\star k} \delta_u (T_j^{(\alpha_1, \cdots,\alpha_j)})$ are local.
\hfill $\diamond$

In general, the symmetry $Q_0$ defined in Proposition \ref{pro5} is not a symmetry of
operator $\Re_2$, that is, $L_{Q_0} \Re_2 \neq 0$. Therefore, $[Q_0, Q_1]\neq 0$.
Although $\Re_2$ is a Nijenhuis operator, the generated symmetries do not commute.
Furthermore, $\Re_2^{\star} \delta_u (T_j^{(\alpha_1, \cdots,\alpha_j)})$ are no
longer closed.

Let us look at a simple case when $j=1$. The corresponding cosymmetry is
\begin{eqnarray*}
&& \delta_u T_1^{(\alpha)}=\delta_u (u_1^{2-4 \alpha} u_2^{\alpha})= (\alpha-1)
D_x\left( \alpha u_1^{2-4 \alpha} u_2^{\alpha-2} u_3+(2-4 \alpha) u_1^{1-4 \alpha}
u_2^{\alpha}\right) \ .
\end{eqnarray*}
Notice that
\begin{eqnarray*} &&\delta_u (u_1^{-2} u_2 \ln (u_1^{-4} u_2)= D_x\left(
u_1^{-2} u_2^{-1} u_3-2 u_1^{-3} u_2\right) \ .
\end{eqnarray*}
Without losing generality, we let $\Re_2^{\star}$ act on the cosymmetry $$ \xi^{(1)}=
D_x\left( \alpha u_1^{2-4 \alpha} u_2^{\alpha-2} u_3+(2-4 \alpha) u_1^{1-4 \alpha}
u_2^{\alpha}\right),$$ where $\alpha\in \mathbb{C}$ and $\xi^{(1)}$ is closed. We
have
\begin{eqnarray*}
\Re_2^{\star} \xi^{(1)} &=&2 \alpha D_x \left(  u_1^{6-4 \alpha}u_2^{\alpha-4} u_5
+(3 \alpha -7) u_1^{6-4 \alpha}u_2^{\alpha-5}u_3 u_4
+(8-12  \alpha) u_1^{5-4 \alpha}u_2^{\alpha-3} u_4 \right.\\
&&\left.+(\alpha-4)(\alpha-2) u_1^{6-4 \alpha}u_2^{\alpha-6} u_3^3
-4 (3 \alpha-2) (\alpha-2) u_1^{5-4 \alpha}u_2^{\alpha-4} u_3^2\right.\\
&&\left. +2 (4 \alpha-1) (6 \alpha-5) u_1^{4-4 \alpha}u_2^{\alpha-2} u_3 -16 (4
\alpha-1) (\alpha-1) u_1^{3-4 \alpha}u_2^{\alpha} \right),
\end{eqnarray*}
which is a cosymmetry of equation (\ref{HSn}), that is, $L_{u_t} (\Re_2^{\star}
\xi^{(1)})=0$. As we mentioned in Section \ref{Sec2}, ${\rm d} (\Re_2^{\star}
\xi^{(1)})\neq 0$ is a symplectic operator. We have the following.
\begin{Pro}\label{pro6}
Operator $ \J=D_x (D_x +\frac{u_3}{u_2}) \left(\frac{4 u_2^2- u_1 u_3} {u_1^{4
\alpha-5} u_2^{5-\alpha}}D_x +D_x \frac{4 u_2^2- u_1 u_3}{u_1^{4 \alpha-5}
u_2^{5-\alpha}} \right) (D_x -\frac{u_3}{u_2}) D_x$ is a symplectic operator for the
Hunter-Saxton equation (\ref{HSn}) for all $\alpha\in \mathbb{C}$.
\end{Pro}
{\bf Proof}. By direct calculation, we have
\begin{eqnarray}
&&{\rm d} (\Re_2^{\star} \xi^{(1)})=D_{\Re_2^{\star} \xi^{(1)}}-D_{\Re_2^{\star}
\xi^{(1)}}^{\star}\nonumber\\
&=&(\alpha+1) D_x (D_x +\frac{u_3}{u_2}) \left(\frac{4
u_2^2- u_1 u_3} {u_1^{4 \alpha-5} u_2^{5-\alpha}}D_x +D_x \frac{4 u_2^2- u_1
u_3}{u_1^{4 \alpha-5} u_2^{5-\alpha}} \right) (D_x -\frac{u_3}{u_2}) D_x\ ,\label{dcos}
\end{eqnarray}
which is a symplectic operator when $\alpha\neq -1$. Since the Lie derivative commutes
with ${\rm d}$, that is
$$L_{u_t} {\rm d} (\Re_2^{\star} \xi^{(1)})={\rm d} L_{u_t}(\Re_2^{\star} \xi^{(1)})=0$$
implying that it is a symplectic operator for the Hunter-Saxton equation
(\ref{HSn}). When $\alpha=-1$, we can write
\begin{eqnarray*}
 \J={\rm d} (D_x S).
\end{eqnarray*}
Here $S=\frac{u_1^{10} u_5}{u_2^5}-9 \frac{u_1^{10} u_3 u_4}{u_2^6}
+\frac{25}{2}\frac{u_1^{10} u_3^3}{u_2^7} +16 \frac{u_1^{9} u_4}{u_2^4}-43
\frac{u_1^{9} u_3^2}{u_2^5}+66 \frac{u_1^{8} u_3}{u_2^3}-96 \frac{u_1^{7} }{u_2}$ is
a symmetry of the Hunter-Saxton equation (\ref{HSn}), which can be verified according
to Definition \ref{defs}. By the same reason as above, we prove that $\J$ is a
symplectic operator for the Hunter-Saxton equation. \hfill $\diamond$

For formula (\ref{dcos}) when $\alpha=-1$, we have ${\rm d} (\Re_2^{\star}
\xi^{(1)})=0$, which is consistent with the result in Theorem \ref{th6}. By the
Leibniz rule for the Lie derivative, any product of Hamiltonian and symplectic operators
of the Hunter-Saxton equation is a recursion operator. Using its Hamiltonian operator
$D_x^{-1}$ and a symplectic operator $\J$ as in Proposition \ref{pro6}, we obtain the
following result:
\begin{Cor}\label{cor2}
Differential operator $(D_x +\frac{u_3}{u_2}) \left(\frac{4 u_2^2- u_1 u_3} {u_1^{4
\alpha-5} u_2^{5-\alpha}}D_x +D_x \frac{4 u_2^2- u_1 u_3}{u_1^{4 \alpha-5}
u_2^{5-\alpha}} \right) (D_x -\frac{u_3}{u_2}) D_x$ is a recursion operator for the
Hunter-Saxton equation (\ref{HSn}) for $\alpha\in \mathbb{C}$.
\end{Cor}
Here we can use different Hamiltonian operators in Theorem \ref{th3} instead of
$D_x^{-1}$. However, the other Hamiltonian operators will bring in nonlocal terms in
the recursion operators. Further study is required to determine whether the resulting
operators produce local symmetries or not.
\section{Discussion}\label{Sec5}
It is well known that the integrable equations possess an infinite number 
of commuting conserved densities and generalised symmetries. 
In this paper, we present a new feature for the integrable Hunter-Saxton equation: infinitely many noncommuting $x,t$-independent conserved densities and symmetries.
We found three Nijenhuis recursion operators and a
local parameter-dependent recursion operator. We believe that there are more
Nijenhuis recursion operators related to the conserved densities listed in Theorem
\ref{th2} since we have found a new commuting pair (\ref{newc}) in Appendix A. Note that
we can define the Poisson bracket (\ref{poid}) with respect to $\cH_2$ instead of
$D_x^{-1}$. It will be interesting to extend the study in the paper and to
see whether there are new commuting pairs.

The conserved densities in Theorem \ref{th2} give rise to infinitely many
cosymmetries of the Hunter-Saxton equation (\ref{HSn}), which are closed $1$-forms.
In Section \ref{sec42}, we showed that the results of recursion operator $\Re_2$ (\ref{r2}) acting
on such closed cosymmetries are no longer closed. Hence, we can generate a lot of
local symplectic operators such as in Proposition \ref{pro6}. This will lead to local
recursion operators as in Corollary \ref{cor2}. The immediate questions are: what are the
relations among such recursion operators? Can we write down neat formulas for them?
Using the computer algebra system \maple\ , it is not hard to compute these operators
although the expressions are huge. The problem is to present them in a compact way, e.g.
as the product of $1^{\rm st}$ order differential operators as in Proposition
\ref{pro6}.

We know the set of symmetries is a Lie algebra under Lie bracket (\ref{lie}). For noncommuting
symmetries of the Hunter-Saxton equation, we can use them to generate higher order symmetries. They are different from master symmetries  \cite{mr86c:58158}, which generate commuting symmetries.

The fundamental question is: where do such rich structures for the Hunter-Saxton
equation come from? As mentioned in the beginning of the paper, the equation is
linearizable and has a trivial dispersion law. However, the transformation
(\ref{tran}) is highly nonlocal. We didn't find a direct way to produce the present
results from the linearized equation (\ref{linhs}). It would be very helpful for
getting a clear and complete picture if we could find the direct link.
\section*{Appendix A}\label{A}
In this Appendix, we give the \maple\ result of the commutator of conserved densities
between $T_1^{(\alpha)}$ and $T_3^{(\beta,\gamma,\mu)}$. We also list the new
commuting pair between $ T_2^{(\alpha_1, \alpha_2)}$ and $
T_3^{(\beta_1,\beta_2,\beta_3)}$.

Using the computer algebra system \maple\ , we obtain the following formula:
\begin{eqnarray*}
&& \left\{ T_1^{(\alpha)},\  T_3^{(\beta,\gamma,\mu)}\right\}_{D_x^{-1}} =\frac{1}{2} \alpha
(\alpha-1) \mu (\mu-1) (2-\mu) \ T_4^{(\alpha+\beta-2,\gamma,\mu-3,3)}\\
&&\qquad -\frac{3}{2} \alpha (\alpha-1) \mu (\mu-1) \gamma \
T_4^{(\alpha+\beta-2,\gamma-1,\mu-1,2)}
\\
&&\qquad -\frac{1}{2} \alpha (\alpha-1) \mu (\mu-1)(3 \beta-7 \alpha+14) \
T_4^{(\alpha+\beta-3,\gamma+1,\mu-2,2)}
\\
&&\qquad + \alpha (\alpha-1) \gamma (\gamma-1)(\mu-1) (\mu+1) \
T_4^{(\alpha+\beta-2,\gamma-2,\mu+1,1)}
\\
&&\qquad + \alpha (\alpha-1) (2 \mu^2 \beta \gamma-\beta \gamma \mu+\mu^2 \gamma^2
\alpha-4 \alpha \gamma \mu^2+8 \mu^2 \gamma
+3 \alpha \gamma \mu -2 \gamma^2 \mu^2-6 \gamma \mu \\
&&\qquad  \qquad +\mu^2 \beta +6 \mu +\alpha \gamma +\mu \beta +2 \gamma^2-6 \mu^2+3
\mu^2 \alpha -\alpha \gamma^2 -2 \gamma -3 \mu \alpha) \
T_4^{(\alpha+\beta-3,\gamma,\mu,1)}\\
&&\qquad + \alpha \gamma (\alpha-1) (\gamma-1) (\gamma-2) (\mu -1) \
T_3^{(\alpha+\beta-2,\gamma-3, \mu+3)}\\
&&\qquad + \alpha \gamma (\alpha-1) (-2 \gamma^2 \mu-\alpha \gamma^2+\alpha \gamma^2
\mu +2 \gamma^2 +3 \gamma \beta \mu -6 \gamma +3 \alpha \gamma -2 \beta \gamma +6
\gamma \mu
\\ &&\qquad \qquad  -3 \alpha \gamma \mu +2 \alpha \mu -2 \alpha +2 \beta -4 \mu +4)
\ T_3^{(\alpha+\beta-3,\gamma-1, \mu+2)}\\
&&\qquad + \alpha (\alpha-1) (72+2 \gamma \beta \mu+13 \beta +24 \gamma
-72 \mu -60 \alpha +4 \alpha^2 \gamma  +\beta^2 \mu+2 \beta^2 \gamma \mu\\
 &&\qquad \qquad  +\alpha^2 \gamma^2 \mu -4 \alpha^2 \gamma \mu -\alpha^2 \gamma^2+\alpha \beta
\gamma^2 \mu +4 \alpha \beta \mu -\alpha \beta \gamma^2 -8 \alpha \beta -12 \alpha^2 \mu\\
 &&\qquad \qquad  -2 \alpha \beta \gamma \mu +60 \alpha \mu -9 \beta \mu +3 \beta^2 +12 \alpha^2 -6
\gamma^2 -24 \gamma \mu +2 \beta \gamma^2 +5 \alpha \gamma^2\\
 &&\qquad \qquad   -2 \beta \gamma^2 \mu +20 \alpha \gamma \mu -20 \alpha \gamma +6 \gamma^2 \mu -5
\alpha \gamma ^2 \mu ) \ T_3^{(\alpha+\beta-4,\gamma+1, \mu+1)}\\
&&\qquad + \alpha (\alpha-1) (240 \mu -10 \mu \alpha^3-8 \alpha \beta -54 \alpha^2
-144 +24 \gamma +9 \alpha^2 \gamma
+10 \beta +\beta^3+\beta^2 \\
&&\qquad \qquad   +156 \alpha-260 \alpha \mu +2 \alpha^2 \beta +6 \alpha^3-2 \alpha
\beta^2 +90 \mu \alpha^2 -26 \alpha \gamma -\gamma \alpha^3) \
T_3^{(\alpha+\beta-5,\gamma+3,\mu)}\\
&&\qquad + \alpha (\alpha-1) (\alpha-2) (\alpha-3) (\alpha-4) (5-\alpha) \mu \
T_3^{(\alpha+\beta-6,\gamma+5,\mu-1)}
\end{eqnarray*}
Combining with formula (\ref{t12}), we can obtain the following result:
\begin{Pro}\label{pro7}
Assume that $\alpha(\alpha-1) \mu (\mu-1)\neq 0$. There are only three
commuting pairs among $T_1^{(\alpha)}$ and  the third order conserved densities, namely,
\begin{eqnarray*}
&&\left\{ T_1^{(1/2)},\  T_3^{(-7/2,0, 2)}-\frac{35}{16} T_2^{(-11/2,4)}\right\}_{D_x^{-1}}
=\left\{ T_1^{(-1)},\  T_3^{(-7,0, 2)}-7\ T_2^{(-9,4)}\right\}_{D_x^{-1}}\\
&&=\left\{ T_1^{(2)},\  T_3^{(0,0, 2)}\right\}_{D_x^{-1}}=0\ .
\end{eqnarray*}
\end{Pro}
{\bf Proof.} We search for the linear combinations of
$T_3^{(\beta,\gamma,\mu)}+\alpha_1 T_2^{(\beta_1,\gamma_1)}$, which are in involution
of $T_1^{(\alpha)}$ under the Poisson bracket (\ref{poid}). From Theorem \ref{th2},
we know that
\begin{eqnarray*}
&&T_3^{(\nu_1,\nu_2,1)}\equiv
\frac{-\nu_1}{\nu_2+1}T_2^{(\nu_1-1,\nu_2+2)};\\
&&T_4^{(\nu_1,\nu_2,\nu_3,1)}\equiv
-\frac{\nu_1}{\nu_3+1}T_3^{(\nu_1-1,\nu_2+1,\nu_3+1)}
-\frac{\nu_2}{\nu_3+1}T_3^{(\nu_1,\nu_2-1,\nu_3+2)}\ .
\end{eqnarray*}
We substitute the above formula into $\left\{ T_1^{(\alpha)},\
T_3^{(\beta,\gamma,\mu)}+\alpha_1 T_2^{(\beta_1,\gamma_1)}\right\}_{D_x^{-1}}$ and
write it as the combination of independent terms. Note that it vanishes if and only
if all coefficients of independent terms are equal to zero. We immediately get
$\mu=2$ and $\gamma=0$. The other conditions are
\begin{eqnarray*}
&& \left\{ \begin{array}{l} 7 \alpha-3 \beta-14=0;\\
\beta=\beta_1+2;\\
\gamma_1=4;\\
(-84+70 \alpha   +3 \beta^2  -4 \alpha \beta  +5 \beta   -14 \alpha^2  )
+2 \alpha_1  (\gamma_1-1) (2-\gamma_1)=0;\\
336 -14 \alpha^3-8 \alpha \beta +10 \beta +\beta^3+\beta^2  -364 \alpha  +2 \alpha^2 \beta
-2 \alpha \beta^2 +126 \alpha^2\\
\qquad \qquad \qquad + 6 \alpha_1   (5 \alpha-3 \beta_1-10)=0;\\
3 \alpha_1   (\alpha-2)  \beta_1  (1-\beta_1)-\frac{1}{6}(\alpha+\beta_1-4)
\left(2 (\alpha-2) (\alpha-3) (\alpha-4) (5-\alpha)\right.\\
\left.\qquad \qquad \qquad+3 \alpha_1   (24+4 \alpha^2+\beta_1
-4 \alpha \beta_1  +8 \beta_1-\beta_1^2-20 \alpha)\right)=0\
.\end{array}\right.
\end{eqnarray*}
Solving this algebraic system, we obtain the following three solutions
\begin{eqnarray*}
 \left\{ \begin{array}{l}\alpha=\frac{1}{2} \\ \alpha_1=-\frac{35}{16}\\ \beta_1=-\frac{11}{2}\\
\gamma_1=4\\ \beta=-\frac{7}{2}
        \end{array}\right. \qquad
 \left\{ \begin{array}{l}\alpha=-1 \\ \alpha_1=-7\\ \beta_1=-9\\
\gamma_1=4\\ \beta=-7
        \end{array}\right. \qquad
\left\{ \begin{array}{l}\alpha=2\\ \alpha_1=0\\ \beta_1=-2\\
\gamma_1=4\\ \beta=0
        \end{array}\right.
\end{eqnarray*}
These correspond to the commuting pairs listed in the statement.
\hfill $\diamond$

Notice that we get the same values of $\alpha$ as in Proposition \ref{pro3}. In
section \ref{sec41}, we show that there are three Nijenhuis recursion operators
corresponding to each value of $\alpha$. These three commuting pairs can be directly
found from the corresponding recursion operators. For example,
\begin{eqnarray*}
\Re_2^{\star} (\delta_u T_2^{(-5, 2)})=
- 2 \delta_u \left(T_3^{(-7,0, 2)}-7\ T_2^{(-9,4)}\right)\ .
\end{eqnarray*}
Proposition \ref{pro7}  implies that there are no other commuting pairs between
$T_1^{(\alpha)}$ and conserved densities generated in Theorem \ref{th2} of third
order. We conjecture that there are only three commuting pairs between
$T_1^{(\alpha)}$ and conserved densities generated in Theorem \ref{th2} of any higher
order.

We compute the commutator between $T_2^{(\alpha_1, \alpha_2)}$ and $T_2^{(\alpha_1,
\alpha_2, \alpha_3)}$. Beside the three pairs directly obtained from Proposition
\ref{pro3} and Proposition \ref{pro7}, we also find the following new commuting pair:
\begin{eqnarray}\label{newc}
\left\{ T_2^{(0,1/3)}\!\!\!,\  T_3^{(0,-7/3, 2)}\right\}_{D_x^{-1}}\!\!\!=
\left\{\left(\frac{u_1 u_3-4 u_2^2}{u_1}\right)^{1/3}\!\!\!,\
\frac{(u_1^2 u_4-14 u_1 u_2 u_3 +28 u_2^3)^2}{u_1^{5/3}
\left(u_1 u_3-4 u_2^2\right)^{7/3}} \right\}_{D_x^{-1}}\!\!\!\!=0.
\end{eqnarray}
We have not found the corresponding Nijenhuis recursion operator for this new case as
we did in section \ref{sec41}.

\section*{Appendix B: Proof of Theorem \ref{th4}}\label{B}
In this Appendix, we give the proof of Theorem \ref{th4}. Let us first introduce some
notation:
$$f_i=D_x^{i} f, \qquad  f^{(j)}=\frac{\partial f}{\partial u_j},
\qquad (f \theta)_{-1}=D_x^{-1} (f \theta).$$
The same notation is also used for $g$ and $h$.

{\bf Proof}. To prove the statement, we check when the operator
\begin{eqnarray*}
&&{\cal H}=f(u_1,u_2) D_x+D_x f +g(u_1,u_2,u_3) D_x^{-1} h(u_1,u_2,u_3)
+h D_x^{-1} g+\lambda D_x^{-1}
\end{eqnarray*}
is Hamiltonian for arbitrary constant $\lambda\in \mathbb{C}.$  From Chapter 7 in
\cite{mr94g:58260}, We know $\cH$ is Hamiltonian if and only if
\begin{eqnarray*}
 \Pr V_{\cH \theta}(\Theta_{\cH})=0,\quad \mbox{where} \quad
\Theta_{\cH} =\frac{1}{2} \int \left(  \theta \wedge \cH \theta \right) {\rm d} x
\end{eqnarray*}
is the associated bi-vector of \(\cH\).

First we have
\begin{eqnarray*}
&&{\cal H}(\theta)=2 f \theta_1+f_1 \theta+g (h \theta)_{-1}+h (g
\theta)_{-1}+\lambda \theta_{-1};\\
&&D_x({\cal H}(\theta))=2 f \theta_2+3 f_1 \theta_1+f_2 \theta+2 g h \theta +g_1 (h
\theta)_{-1}+h_1 (g \theta)_{-1}+\lambda \theta;\\
&&D_x^2({\cal H}(\theta))=2 f \theta_3+5 f_1 \theta_2+4 f_2 \theta_1+f_3 \theta+2 g h
\theta_1+3 (g h)_1 \theta +g_2 (h \theta)_{-1}+h_2 (g \theta)_{-1}+\lambda \theta_1;\\
&&D_x^3({\cal H}(\theta))=2 f \theta_4+7 f_1 \theta_3+9 f_2 \theta_2+5 f_3
\theta_1+f_4 \theta+2 g h \theta_2+5 (g h)_1 \theta_1 +4 (g h)_2 \theta -2 g_1 h_1
\theta \\&& \qquad\qquad+g_3 (h \theta)_{-1}+h_3 (g \theta)_{-1}+\lambda \theta_2.
\end{eqnarray*}
Substituting them into $ \Pr V_{\cH \theta}(\Theta_{\cH})$,
this leads to
\begin{eqnarray*}
{\rm Pr}_{{\cal H}(\theta)} \Theta_{\cH}&=&\int \left( \theta \wedge D_f[{\cal
H}(\theta)] \wedge \theta_1+\theta \wedge  D_g[{\cal H}(\theta)]\wedge (h
\theta)_{-1}+\theta \wedge  D_h[{\cal H}(\theta)]\wedge (g \theta)_{-1} \right) {\rm d} x\\
&=&\int  \theta \wedge \left(f^{(1)} D_x({\cal H}(\theta))+f^{(2)} D_x^2({\cal
H}(\theta))\right)\wedge \theta_1 {\rm d} x\\
&&+ \int  \theta \wedge \left(g^{(1)} D_x({\cal H}(\theta))+g^{(2)} D_x^2({\cal
H}(\theta))+g^{(3)} D_x^3({\cal H}(\theta))\right)\wedge (h \theta)_{-1} {\rm
d} x\\
&&+ \int \theta \wedge  \left(h^{(1)} D_x({\cal H}(\theta))+h^{(2)} D_x^2({\cal
H}(\theta))+h^{(3)} D_x^3({\cal H}(\theta))\right) \wedge (g \theta)_{-1}  {\rm d} x\ .
\end{eqnarray*}
It needs to vanish for all $\lambda$. So the coefficient of $\lambda$ should be zero, that is
\begin{eqnarray*}
0&=&\int  \theta \wedge (g^{(2)} \theta_1+g^{(3)} \theta_2)\wedge (h \theta)_{-1}
{\rm d} x+ \int \theta \wedge  (h^{(2)} \theta_1+h^{(3)} \theta_2) \wedge (g
\theta)_{-1}  {\rm d} x\\
&=& \int \left( \theta \wedge (g^{(2)} -g^{(3)}_1) \theta_1 \wedge (h \theta)_{-1} +
\theta \wedge  (h^{(2)} -h^{(3)}_1) \theta_1 \wedge (g \theta)_{-1} \right) {\rm d} x\ .
\end{eqnarray*}
This leads to $h^{(2)}=h^{(3)}_1$ and $g^{(2)}=g^{(3)}_1$. The implies that
\begin{eqnarray}\label{con1}
h(u_1,u_2, u_3)=p_h(u_1, u_2) u_3 +q_h (u_1,u_2), \qquad g(u_1,u_2, u_3)=p_g(u_1,
u_2) u_3 +q_g (u_1,u_2)
\end{eqnarray}
and
\begin{eqnarray}\label{con2}
q_h^{(2)}=p_h^{(1)} u_2, \qquad q_g^{(2)}=p_g^{(1)} u_2\ .
\end{eqnarray}
The rest of the terms, i.e., the terms without $\lambda$  in $ \Pr V_{\cH
\theta}(\Theta_{\cH})$ should also vanish.
\begin{eqnarray*}
0&=&\int  f^{(1)} \theta \wedge \left(2 f \theta_2+g_1 (h \theta)_{-1}
+h_1 (g \theta)_{-1}\right)\wedge \theta_1 {\rm d} x\\
&&+\int  f^{(2)} \theta \wedge \left(2 f \theta_3+5 f_1 \theta_2
+g_2 (h \theta)_{-1}+h_2 (g \theta)_{-1}\right)\wedge \theta_1 {\rm d} x\\
&&+ \int g^{(1)} \theta \wedge \left(2 f \theta_2+3 f_1 \theta_1+h_1 (g
\theta)_{-1}\right)\wedge (h \theta)_{-1} {\rm
d} x\\
&&+ \int g^{(2)} \theta \wedge \left(2 f \theta_3+5 f_1 \theta_2+4 f_2 \theta_1+2 g h
\theta_1 +h_2 (g \theta)_{-1}\right)\wedge (h \theta)_{-1} {\rm
d} x\\
&&+ \int\!\! g^{(3)} \theta \wedge \left(2 f \theta_4\!+\!7 f_1 \theta_3\!+\!9 f_2 \theta_2\!+\!5 f_3
\theta_1\!+\!2 g h \theta_2\!+\!5 (g h)_1 \theta_1 +h_3 (g \theta)_{-1}\right)\wedge (h
\theta)_{-1} {\rm
d} x
\end{eqnarray*}
\begin{eqnarray*}
&&+ \int h^{(1)} \theta \wedge  \left(2 f \theta_2+3 f_1 \theta_1+g_1 (h
\theta)_{-1}\right) \wedge (g \theta)_{-1}  {\rm d}
x\\
&&+ \int h^{(2)} \theta \wedge  \left(2 f \theta_3+5 f_1 \theta_2+4 f_2 \theta_1+2 g
h \theta_1 +g_2 (h \theta)_{-1}\right) \wedge (g \theta)_{-1} {\rm d}
x\\
&&+ \int\!\! h^{(3)} \theta \wedge  \left(2 f \theta_4+7 f_1 \theta_3+9 f_2 \theta_2+5
f_3 \theta_1+2 g h \theta_2+5 (g h)_1 \theta_1 +g_3 (h \theta)_{-1}\right) \wedge (g
\theta)_{-1} {\rm d} x
\\
&=& \int \left(2 D_x(f^{(2)} f)-5 f_1 f^{(2)}  -2 f f^{(1)} +2 f h g^{(3)} +2 f g
h^{(3)}\right) \theta \wedge \theta_1 \wedge  \theta_2
 {\rm d} x\\
&&+\int  \left( -D_x^3(2 f  g^{(3)}) + D_x^2(2 f  g^{(2)}+7 f_1 g^{(3)})
+D_x(-2 f  g^{(1)}-5 f_1 g^{(2)} -9 f_2 g^{(3)}-2 g h  g^{(3)})  \right.\\
&& \ \ \left. +4 f_2  g^{(2)}+2 g h  g^{(2)}+5 f_3 g^{(3)}+
5 (g h)_1 g^{(3)}\!\!+\!\!3 f_1  g^{(1)}\!\! -\!g_1 f^{(1)}\!\! -\!g_2 f^{(2)} \right)
\theta \wedge \theta_1 \wedge (h \theta)_{-1}  {\rm d} x\\
&&+\int   \left( -D_x^3(2 f  h^{(3)}) + D_x^2(2 f  h^{(2)}+7 f_1 h^{(3)})
+D_x(-2 f  h^{(1)}-5 f_1 h^{(2)} -9 f_2 h^{(3)}-2 g h  h^{(3)})  \right.\\
&& \ \ \left. +4 f_2  h^{(2)}+2 g h  h^{(2)}+5 f_3 h^{(3)}+5 (g h)_1 h^{(3)}+3 f_1  h^{(1)}
-h_1 f^{(1)} -h_2 f^{(2)} \right) \theta \wedge \theta_1 \wedge (g \theta)_{-1}  {\rm d} x\\
&&+ \int \left( g^{(1)} h_1 +g^{(2)} h_2+g^{(3)} h_3-h^{(1)} g_1
-h^{(2)} g_2-h^{(3)} g_3\right) \theta \wedge
(g \theta)_{-1}\wedge (h \theta)_{-1} {\rm
d} x\\
&&+ \int \left( D_x (4 f g^{(3)}) -7 f_1 g^{(3)}-2 f g^{(2)} \right)
\theta_1 \wedge \theta_2 \wedge (h \theta)_{-1} {\rm
d} x\\
&&+ \int \left( D_x (4 f h^{(3)}) -7 f_1 h^{(3)}-2 f h^{(2)} \right)
\theta_1 \wedge \theta_2 \wedge (g \theta)_{-1} {\rm
d} x \ .
\end{eqnarray*}
This implies that every coefficient should be equal to zero, that is
\begin{eqnarray*}
\left\{\begin{array}{l} 2 D_x(f^{(2)} f)-5 f_1 f^{(2)}  -2 f f^{(1)}
+2 f h g^{(3)} +2 f g h^{(3)}=0;\\
-D_x^3(2 f  g^{(3)}) + D_x^2(2 f  g^{(2)}+7 f_1 g^{(3)})+D_x(-2 f  g^{(1)}
-5 f_1 g^{(2)} -9 f_2 g^{(3)}-2 g h  g^{(3)}) \\
\qquad  +4 f_2  g^{(2)}+2 g h  g^{(2)}+5 f_3 g^{(3)}+5 (g h)_1 g^{(3)}
+3 f_1  g^{(1)} -g_1 f^{(1)} -g_2 f^{(2)}=0;\\
 -D_x^3(2 f  h^{(3)}) + D_x^2(2 f  h^{(2)}+7 f_1 h^{(3)})
 +D_x(-2 f  h^{(1)}-5 f_1 h^{(2)} -9 f_2 h^{(3)}-2 g h  h^{(3)}) \\
\qquad +4 f_2  h^{(2)}+2 g h  h^{(2)}+5 f_3 h^{(3)}+5 (g h)_1 h^{(3)}+3 f_1 h^{(1)}
-h_1 f^{(1)} -h_2 f^{(2)}=0;\\
g^{(1)} h_1+g^{(2)} h_2+g^{(3)} h_3-h^{(1)} g_1-h^{(2)} g_2-h^{(3)} g_3=0;\\
4 f g^{(3)}) -7 f_1 g^{(3)}-2 f g^{(2)}=0;\\
4 f h^{(3)}) -7 f_1 h^{(3)}-2 f h^{(2)}=0\ .
\end{array}\right.
\end{eqnarray*}
Substituting (\ref{con1}) into the above formulas and combining with (\ref{con2}), we
obtain over-determined partial differential equations for the functions $f(u_1\!,u_2)$, $p_h(u_1\!,u_2)$,  $q_h(u_1\!,u_2)$, $g_h(u_1\!,u_2)$ and $q_h(u_1\!,u_2)$. With
the help of the package {\it diffalg} in {\rm Maple}, we obtain the five cases listed in
Theorem \ref{th4}. \hfill $\diamond$

\section*{Acknowledgements}
The author would like to thank D.D. Holm, A.N.W. Hone, A.V. Mikhailov and P.J. Olver
for inspiring discussions and useful comments. The author also thank the organizers
of NEEDS 2009 for the invitation and financial support, where the author began this
research.

\end{document}